\documentclass[preprint,showpacs,preprintnumbers,amsmath,amssymb]{revtex4}

\usepackage{graphicx}% Include figure files
\usepackage{dcolumn}% Align table columns on decimal point
\usepackage{bm}% bold math

\begin{document}
%\alph{footnote}
%\preprint{{hep-th/0512115} \hfill {UCVFC-DF-25-2002}}
\title{Quantization of Interacting Non-Relativistic Open Strings using Extended Objects}
\author{P. J. Arias}
 \email{parias@fisica.ciens.ucv.ve}
 \author{N. Bol\'{\i}var}
 \email{nbolivar@fis-lab.ciens.ucv.ve}
\author{E. Fuenmayor}%
 \email{efuenma@fisica.ciens.ucv.ve}
\author{Lorenzo Leal}
 \email{lleal@fisica.ciens.ucv.ve}
\affiliation{Centro de F\'{\i}sica Te\'orica y Computacional,
Facultad de Ciencias, Universidad Central de Venezuela, AP 47270,
Caracas 1041-A, Venezuela}%

\date{\today}% It is always \today, today,

%\twocolumn
\begin{abstract}

Non-relativistic charged open strings coupled with Abelian gauge
fields are quantized in a geometric representation that
generalizes the Loop Representation. The model comprises
open-strings interacting through a Kalb-Ramond field in four
dimensions. It is shown that a consistent geometric-representation
can be built using a scheme of ``surfaces and lines of Faraday'',
provided that the coupling constant (the ``charge'' of the string)
is quantized.

\end{abstract}
\pacs{Valid PACS appear here}% PACS, the Physics and Astronomy

\maketitle

%\begin{multicols}2
\section{Introduction}

In this paper we consider theories of interacting non-relativistic
strings, and quantize them in a representation that uses extended
geometrical objets (paths and surfaces), that generalize the usual
Loop Representation ($LR$) \cite{GT}. Special attention is devoted
to the geometric quantization of interacting open strings. The
surface representation was considered years ago to study the
free-field case \cite{PIO,Pio}, but it has to be adapted to
include the particularities that the coupling with the string
demands. To obtain a proper understanding of this theory we first
consider the model of closed strings in self-interaction by means
of an Abelian Kalb-Ramond field \cite{KR}. This model can be seen
as a generalization of the theory of charged non-relativistic
point particles, in electromagnetic interaction, quantized within
the $LR$, exposed in reference \cite{E1}, where it was found that
the charge must be quantized in order to the $LR$ formulation
(Faraday's lines scheme) of the model be consistent. This result
agrees with those obtained in previous developments
\cite{Corichi1,Corichi2}. For both theories (interaction of closed
and open strings) we find that the coupling constant of the string
(let us say, the Kalb-Ramond ``charge'' of the string) must be
also quantized, if the geometric representation adapted to the
model is going to be plausible. On other hand, we should be aware
that when we couple extended material objects to fields, the
theory presents certain subtleties regarding its quantization, and
so does the appropriate geometric representation.

The next section is devoted to the study of a geometric open
path-surface representation for non-relativistic ``charged''
strings interacting by means of the Kalb-Ramond field. We discuss
first the case of closed strings (subsection \ref{A}) treated
previously in reference \cite{E} and in subsection \ref{B} we
consider the more interesting open string model, in a geometric
representation. Final remarks are left to the last section.

\section{Non-relativistic interacting ``charged'' strings}\label{sec2}

\subsection{Closed string interaction}\label{A}

Let us start our discussion by introducing the action \cite{E}
\begin{equation}\label{2.1}
S=\frac{1}{12{g^2}}\int H^{\mu\nu\lambda}H_{\mu\nu\lambda}d^{4}x +
\frac{\alpha}{2}\int dt \int d\sigma \left[(\dot{z}^i)^2-(z'^i)^2)
\right] + \frac{1}{2} \int d^4x
\textit{J}^{\mu\nu}B_{\mu\nu},\nonumber\\
\end{equation}
where the Kalb-Ramond antisymmetric potential and field strength,
$B_{\mu\nu}$ and $H_{\mu\nu\lambda}$, respectively, are related by
$H_{\mu\nu\lambda}=3\partial_{[\mu}B_{\nu\lambda]}=\partial_{\mu}
B_{\nu\lambda}+\partial_{\lambda} B_{\mu\nu}+\partial_{\nu}
B_{\lambda\nu}$. Also we have a contribution corresponding to the
free non-relativistic closed string, whose world sheet spatial
coordinates $z^{i}(t,\sigma)$ are given in terms of the time $t$
and the parameter $\sigma$ along the string. The string tension
$\alpha$ has units of $mass^2$ and $g$ is a parameter with units
of $mass$. The string-field interaction term is given by means of
the current $\textit{J}^{\mu\nu}(\vec{x},t)= \phi \int dt \int
d\sigma \left[\dot{z}^\mu z'^\nu -\dot{z}^\nu
z'^\mu\right]\delta^{(4)}(x-z)$. Here $\phi$ is the dimensionless
coupling constant (analog to the charge in the case of particles),
and we indicate with dots and primes partial derivation with
respect to the time parameter $t$ and $\sigma$, respectively.

We are interested in the Dirac quantization scheme of the theory
so we define the conjugate momenta associated to the fields,
$B_{ij}$, and string variables, $z^i$, as
\begin{equation}\label{2.2}
\Pi^{ij}=\frac{1}{2g^2}\left(\dot{B}_{ij}+ \partial_j
B_{0i}-\partial_i B_{0j}\right), \quad \quad P_{i}=\alpha
\dot{z}^i +\phi B_{ij}z'^{j},
\end{equation}
and directly obtain the Hamiltonian performing a Legendre
transformation in the variables $B_{ij}$ and $z^{i}$,
\begin{equation}\label{2.3}
H=\int d^3x \left[{g}^2
\Pi^{ij}\Pi^{ij}+\frac{1}{12g^2}H_{ijk}H_{ijk}\right]+\int d\sigma
\frac{\alpha}{2}\left[\frac{1}{\alpha^2}\left(P_{i}-\phi
B_{ij}(z)z'^{j}\right)^{2}+(z'^{i})^2 \right]+\int d^3x
B_{0i}\chi^i .
\end{equation}
$B_{i0}$ are not dynamical variables, they appear as Lagrange
multipliers enforcing the first class constraints
\begin{equation}\label{2.4}
\chi^{i}(x) \equiv - \rho^{i}(x) - 2\partial_j \Pi^{ji}(x)=0,
\end{equation}
where $\rho^{i}(x)\equiv \phi \int d\sigma
z'^i\delta^{(3)}(\vec{x}-\vec{z})$ is the ``charge density'' of
the string. The preservation of the above constraints can be done
using the defined canonical commutators algebra of the operators
involved, whose non-vanishing conmutators are given by
\begin{eqnarray}\label{2.5}
\left[z^i(\sigma),P_j(\sigma')\right]=i\delta^i_j\delta(\sigma
-\sigma')\qquad
\left[B_{ij}(\vec{x}),\Pi^{kl}(\vec{y})\right]=i\frac12\left(\delta_{i}^{k}\delta_{j}^{l}
-\delta_{i}^{l} \delta_{j}^{k}\right)
\delta^{(3)}(\vec{x}-\vec{y}),
\end{eqnarray}
This preservation does not produce new constraints.

Now, in order to solve relation (\ref{2.4}), we introduce a
geometric representation based on extended objects: an
``open-surface representation'', related with the $LR$ formulated
by Gambini and Tr\'{\i}as \cite{GT} (and with a geometrical
formulation based on closed surfaces \cite{PIO,Pio}). Consider the
space of piecewise smooth oriented surfaces (for our purposes) in
$R^{3}$. A typical element of this space, let us say $\Sigma$,
will be the union of several surfaces, perhaps some of them being
closed. In the space  of smoth oriented surfaces $\Sigma$ we
define equivalence classes of surfaces that share the same ``form
factor'' $T^{ij}(x,\Sigma)=\int
d\Sigma^{ij}_y\,\delta^{(3)}(\vec{x}-\vec{y})$, where
$d\Sigma^{ij}_y=(\frac{\partial y^i }{\partial s}\frac{\partial
y^j}{\partial r} - \frac{\partial y^i}{\partial r}\frac{\partial
y^j}{\partial s})dsdr$ is the surface element and $s$, $r$ are the
parametrization variables. All the features of the ``open surfaces
space'', are more or less immediate generalizations of aspects
already present in the Abelian path space \cite{GT,C,LO,E}. Our
Hilbert space is composed by functionals $\Psi(\Sigma)$ depending
on equivalence classes $\Sigma$. We need to introduce the surface
derivative $\delta_{ij}(x)$ defined by,
\begin{equation}\label{2.6}
\Psi (\delta\Sigma\cdot\Sigma)-\Psi
(\Sigma)=\sigma^{ij}\delta_{ij}(x)\Psi (\Sigma),
\end{equation}
that measures the response of $\Psi(\Sigma)$ when an element of
surface whose infinitesimal area $\sigma_{ij}=u^iv^j-v^ju^i$,
generated by the infinitesimal vectors $\vec{u}$ and $\vec{v}$, is
attached to $\Sigma$ at the point $x$ \cite{PIO,Pio,E}.

It can be seen that the fundamental commutator associated to
equation (\ref{2.5}) can be realized on surface-dependent
functionals if one sets
\begin{eqnarray}\label{2.7}
\hat{\Pi}^{ij}(\vec{x})\;\longrightarrow\;
\frac{1}{2}T^{ij}(\vec{x},\Sigma),\qquad
\hat{B}_{ij}(\vec{x})\;\longrightarrow \;
2i\delta_{ij}(\vec{x}),\\
\hat{z}^i(\sigma)\; \longrightarrow\; z^i(\sigma), \qquad
\hat{P}^i(\sigma)\; \longrightarrow\; -i \frac{\delta} {\delta
z^i(\sigma)},
\end{eqnarray}
and then the states of the interacting theory can be taken as
functionals $\Psi[\Sigma,z(\sigma)]$, where the field is
represented by the surface $\Sigma$ and matter by means of the
coordinates of the string world sheet. Of all of these functionals
we must pick out those that belong to the kernel of the Gauss
constraint (\ref{2.4}), now written as
\begin{eqnarray}\label{2.8}
\left(\phi\int_{string} d\sigma z'^i\delta^{(3)}(\vec{x}-\vec{z})
-\int_{\partial\Sigma}d\sigma z'^i
\delta^{(3)}(\vec{x}-\vec{z})\right)\Psi[\Sigma ,z(\sigma)]
\approx 0.
\end{eqnarray}
In the last equation we have used that $\partial_j
T^{ji}(\vec{x},\Sigma)=-T^i(\vec{x},\partial\Sigma)
=-\int_{\partial\Sigma} dz^i \delta^{(3)}(\vec{x}-\vec{z})$, with
$\partial \Sigma$ being the boundary of the surface. At this
point, we recall the geometrical setting that permits to solve the
Gauss constraint in the theory of self-interacting
non-relativistic particles coupled through Maxwell field
\cite{E1}. In that case the physical space may be labelled by
Faraday's lines and the scheme of quantization allows us to
associate to every particle a bundle of lines emanating from or
arriving to it, depending on the sign of the particle's charge.
Then the charge must be quantized, since the number of open paths
(that must be equal to the charge to which they are attached) has
to be an integer \cite{E1}. In the present case, we copy the
interpretation and see that if the surface is such that its
boundary coincides with the string, the constraint (\ref{2.8})
reduces to $\left(\phi-1\right)\,\int_{string}d\sigma z'^j
\delta^{(3)}(\vec{x}-\vec{z})=0$, and it is satisfied in general
for $\phi =1$, so we can say in analogy that the surface emanates
from the string. But it could happen that instead the boundary of
the surface and the string have opposite orientations; in that
case the constraint would be satisfied if $\phi =-1$, and we say
that the surface ``enters'' or ``arrives'' at the string position.
There is also the possibility that the surface could be composed
by several layers ($n$ of them) that start (or end) at the string.
Equation (\ref{2.8}) becomes
$\left(\phi-n\right)\,\int_{string}d\sigma z'^j
\delta^{(3)}(\vec{x}-\vec{z})=0$, and in this case the coupling
constant (``charge'' of the string) must obey $\phi =n$ (the sign
of $n$ depends on the fact that the surfaces  may ``emanate'' from
or ``arrive'' to the source). This is what we call a
representation of ``Faraday's surfaces'' for the
string-Kalb-Ramond system.

\subsection{Open string interaction}\label{B}

Our starting point will be the action,
\begin{eqnarray}\label{1}
  \mathcal{S}=\int dt \int d\sigma \frac{\alpha}{2}\lbrack (\dot{z}^{i})^{2}-(z'^{i})^{2}\rbrack+
  \int d^{4}x \left(\frac{1}{12g^{2}}H^{\mu\nu\lambda}H_{\mu\nu\lambda}-\frac{m^{2}}{4}a^{\mu\nu}a_{\mu\nu}+
  \frac{1}{2}J^{\mu\nu}B_{\mu\nu}+J^{\mu}A_{\mu}\right),\nonumber\\
\end{eqnarray}
where we have defined the $2$-form
$a_{\mu\nu}=B_{\mu\nu}+F_{\mu\nu}$ with
$F_{\mu\nu}=\partial_{\mu}A_{\nu}-\partial_{\nu}A_{\mu}$ inspired
in the St\"{u}ckelberg gauge invariant version of the Proca model
\cite{C}. The vector field $A_{\mu}$ is dimensionless. The
Kalb-Ramond antisymmetric potential and field strength are related
as in (\ref{2.1}). This action is invariant under the simultaneous
gauge transformations,
\begin{eqnarray}
B_{\mu\nu}\; \longrightarrow \; B_{\mu\nu}+\partial_{\mu}
\Lambda_{\nu} -
\partial_{\nu} \Lambda _{\mu}\, ,\qquad\quad
A_{\mu}\; \longrightarrow\;
A_{\mu}-\Lambda_{\mu}+\partial_{\mu}\lambda\, ,\label{3}
\end{eqnarray}
if the currents associated to the matter source (the body and the
extremes of the string) satisfy the relation
\begin{eqnarray}
\partial_{\mu}J^{\mu\nu}+J^{\nu}=0, \label{4}
\end{eqnarray}
which emerges as a consequence of the gauge invariance and
equations of motion. This implies that the current associated to
particles placed in the extremes of the string is conserved
$\partial_{\nu}J^{\nu}=0$, although the string-current
$J^{\mu\nu}$ is not. This is a consequence of the string
interaction term by means of two types of currents associated to
matter, one that is associated to the ``body'' and the other to
the end points. This fact constitutes a desired aspect of the
theory because permits the treatment of open-strings without
losing gauge invariance. The equations of motion are given by
\begin{eqnarray}
\partial_{\mu}H^{\mu\nu\lambda}+m^{2}a^{\nu\lambda}-J^{\nu\lambda}=
0\;,\qquad\quad
m^{2}\partial_{\nu}a^{\nu\lambda}+J^{\lambda}=0,\label{7}
\end{eqnarray}
and they guarantee that (\ref{4}) is satisfied.

Now, we proceed with Dirac quantization procedure. We take
$A_{i}$, $B_{ij}$ and $z^{i}$ as dynamical variables and
$\Pi^{i}$, $\Pi^{ij}$ and $P^{i}$ as their canonical conjugate
momenta, respectively. The Hamiltonian density after a Legendre
transformation in the dynamical variables results to be
\begin{eqnarray}\label{8}
\mathcal{H}&=&g^{2}\Pi^{ij}\Pi^{ij}+\frac{\Pi^{i}\Pi^{i}}{2m^{2}}+\frac{1}{12g^{2}}H_{ijk}H_{ijk}
+\frac{m^{2}}{4}a_{ij}a_{ij}-\frac{1}{2}J^{ij}B_{ij}-J^{i}A_{i}\nonumber
\\&&-B_{0i}(2\partial_{j}\Pi^{ji}+\Pi^{i}+J^{0i})-A_{0}(\partial_{i}\Pi^{i}+J^{0})+\int d\sigma
\frac{\alpha}{2}\left[\frac{1}{\alpha^2}\left(P_{i}-\phi
B_{ij}(z)z'^{j}\right)^{2}+(z'^{i})^2 \right].\nonumber\\
\end{eqnarray}
Just as in the preceding subsection the fields variables $A_{0}$
and $B_{i0}$ are treated as non-dynamical fields from the very
beginning. They appear in $\mathcal{H}$ as Legendre multipliers
enforcing the constraints
\begin{eqnarray}
 \Theta^{i}\equiv2\partial_{j}\Pi^{ji}+\Pi^{i}+J^{0i} \approx 0\, ,\,\qquad
 \Theta\equiv\partial_{j}\Pi^{j}+J^{0}\approx 0, \label{10}
\end{eqnarray}
where it can be seen that they are reducible because
$\partial_{i}\Theta^{i}=\Theta$. Apart from the usual canonical
Poisson algebra between the canonical conjugate variable, it is
straight forward to infere that (see discussion in \cite{C})
\begin{eqnarray}
\{ a_{ij}(\vec{x}), \Pi^{kl}(\vec{y}) \} &=&
\frac{1}{2}(\delta^{k}_{i}\delta^{l}_{j}-\delta^{l}_{i}\delta^{k}_{j})\delta^{(3)}(\vec{x}-\vec{y}), \label{11}\\
\{ a_{ij}(\vec{x}), \Pi^{k}(\vec{y}) \} &=&
(\delta^{k}_{j}\partial_{i}-\delta^{k}_{i}\partial_{j})\delta^{(3)}(\vec{x}-\vec{y}).\label{12}
\end{eqnarray}
It can be shown that the preservation of the constraints
(\ref{10}) does not produce new ones. Furthermore, they result to
be first class constraints that generate time independent gauge
transformations on phase space.

To quantize, the canonical variables are promoted to operators
obeying the commutators that result from the replacement
$\{\;,\;\}\;\longrightarrow\;-i[\;,\;]$. These operators have to
be realized in a Hilbert space of physical states
$|\Psi\rangle_{Phys}$, that obey the constraints (\ref{10}) ($
\Theta^{i}|\Psi\rangle_{Phys}=0$).

At this point, we adapt a geometrical representation to the theory
in terms of extended objects just as it was done in the case of
self interacting closed-strings, but now considering (besides the
surfaces) open paths $\gamma$ associated to the fields $A_{i}$
that mediate the interaction between the extreme points of the
string. We prescribe,
\begin{eqnarray}\label{13}
\hat{\Pi}^{ij}(\vec{x})\;\longrightarrow\;\frac{e}{2}T^{ij}(\vec{x},\Sigma)\,
,\qquad \hat{\Pi}^{i}(\vec{x})\;\longrightarrow\;
eT^{i}(\vec{x},\gamma)\,
,\qquad\hat{a}_{ij}(\vec{x})\;\longrightarrow\;
\frac{2i}{e}\delta_{ij}(\vec{x}),
\end{eqnarray}
where $T^{i}(\vec{x},\gamma)$ is the form factor that describes
the open paths $\gamma$ \cite{GT}. We can see, using
$\partial_{j}T^{ji}(\vec{x},\Sigma)= -T^{i}(\vec{x},\partial
\Sigma)$ and $\delta_{ij}(\vec{x})T^{lk}(\vec{y},\Sigma)=
\frac{1}{2}(\delta^{l}_{i}\delta^{k}_{j}
-\delta^{k}_{i}\delta^{l}_{j})\delta^{(3)}(\vec{x}-\vec{y})$, that
the fundamental commutators associated to equations (\ref{11}) and
(\ref{12}) can be realized when they act over functionals
depending on both surfaces and paths. Also we must pick out those
functionals that belong to the kernel of the constraints
(\ref{10}). In this representation it can be expressed as
\begin{eqnarray}\label{14}
&&\left(e\partial_{i}T^{ij}(\vec{x},\Sigma)+eT^{j}(\vec{x},\partial
\Sigma')+J^{0j}\right) \psi(\Sigma,\gamma,\vec{z}(\sigma))=\nonumber\\
&&\left( -eT^{j}(\vec{x},\partial\Sigma) + eT^{j}(\vec{x},\partial
\Sigma')+\phi\int_{C} d z^{j} \delta^{(3)}(\vec{x}-\vec{z})
\right)\psi(\Sigma,\gamma,\vec{z}(\sigma))\approx 0.
\end{eqnarray}
In (\ref{14}), $\partial \Sigma'$ is the part of the border of the
open surface that is drawn by the open-paths attached to the
extreme points of the open strings. The rest of the surface border
is completed by the strings, i.e., $\partial \Sigma=\partial
\Sigma'+ strings$.

\section{Discussion}\label{sec3}

Henceforth, we have the following interpretation: the states of
the interacting theory of open strings can be taken as functionals
$\Psi[\Sigma, \gamma, z(\sigma)]$ depending on surfaces and paths
(i.e. the equivalence classes discussed above), and functions of
the string variables $z(\sigma)$, that act as a source of the
extended geometrical objects. The body of the string interacts
using a surface, that can, depending on the orientation of the
string (i.e., the coupling constant $\phi$), ``emanate'' or
``arrive'' to it. The end points interact via the open paths (just
as in the case of electromagnetic interaction for particles) that
complete the part $\partial\Sigma'$ of the border of the surface
that is not ``glued'' to the strings. Again, as in the closed
string interaction with the Kalb-Ramond field the surface may
consist of $n$ layers attached to the string (depending on the
value of the coupling constant $\phi$), plus an arbitrary number
of closed surfaces, since the latter do not contribute to the
boundary of the surface.

Following references \cite{E1,E}, one could also consider the
geometric representation of open strings interacting through
topological terms, like a $BF$ term in $3+1$ dimensions. This
study, apart of being interesting regarding the solution of the
constraints when the fields that provide the interaction have a
topological character, has the particularity that the dependence
of the wave-functionals on paths (or more generally, on the
appropriate geometric objects that enter in the representation,
like paths or surfaces) might be eliminated by means of an unitary
transformation \cite{E1,LO}. In that case one could obtain a
quantum mechanics of particles, or particles and strings
(depending on the model), subjected to long range interactions
\cite{E} leading to anomalous statistic
\cite{Leinas,Wilczek,Wu,Indio}. This and other topics shall be the
subject of future investigations.\\

This work was supported by Project $G-2001000712$ of FONACIT.

\end{document}